\documentclass[oldversion]{aa} % double column
\usepackage{graphicx}
\usepackage{color}
\usepackage{longtable}
\usepackage{txfonts}
\usepackage{rotating}
\usepackage{lscape}
\usepackage{natbib}
\newcommand{\gsim}{\;\lower.6ex\hbox{$\sim$}\kern-7.75pt\raise.65ex\hbox{$>$}\;}
\newcommand{\lsim}{\;\lower.6ex\hbox{$\sim$}\kern-7.75pt\raise.65ex\hbox{$<$}\;}

\begin{document}
\title{NGC~6139: a normal massive globular cluster or a first-generation
dominated cluster? Clues from the light elements \thanks{Based on  observations
collected at ESO telescopes under  programme 093.B-0583 and  on data obtained
from the ESO Science Archive Facility under request number 94403}\fnmsep\thanks{
   Tables xx, yy, and zz are only available in electronic form at the CDS via anonymous
   ftp to {\tt cdsarc.u-strasbg.fr} (130.79.128.5) or via
   {\tt http://cdsweb.u-strasbg.fr/cgi-bin/qcat?J/A+A/???/???}}
 }

\author{
A. Bragaglia\inst{1},
E. Carretta\inst{1},
A. Sollima\inst{1}
P. Donati\inst{1,2},
V. D'Orazi\inst{3,4,5},
R.G. Gratton\inst{3},
S. Lucatello\inst{3},
\and
C. Sneden\inst{6}
}

\authorrunning{A. Bragaglia et al.}
\titlerunning{Abundance analysis in NGC 6139}

\offprints{A. Bragaglia, angela.bragaglia@oabo.inaf.it}

\institute{
INAF-Osservatorio Astronomico di Bologna, Via Ranzani 1, I-40127 Bologna, Italy
\and
Dipartimento di Fisica e Astronomia, Universit\`a di Bologna, viale Berti Pichat 6, I-40127 Bologna, Italy
\and
INAF-Osservatorio Astronomico di Padova, Vicolo dell'Osservatorio 5, I-35122
 Padova, Italy
\and
Monash Centre for Astrophysics, School of Physics and Astronomy, Monash University, Melbourne, VIC 3800, Australia
\and
Department of Physics and Astronomy, Macquarie University, Sydney, NSW 2109, Australia
\and
Department of Astronomy and McDonald Observatory, The University of Texas, Austin, TX 78712, USA
  }

\date{}

\abstract{Information on globular clusters (GC) formation mechanisms can be
gathered by studying the chemical signature of the multiple populations that
compose these stellar systems. In particular, we are investigating the
anticorrelations among O, Na, Al, and Mg to explore the influence of cluster
mass and environment on GCs in the Milky Way and in extragalactic systems. We
present here the results obtained on NGC~6139 which, on the basis of its
horizontal branch morphology, had been  proposed to be dominated by
first-generation stars. In our extensive study based on high resolution
spectroscopy, the first for this cluster, we found a metallicity of
[Fe/H]=$-1.579\pm0.015\pm0.058$ (rms=0.040 dex, 45 bona fide member stars) on
the UVES scale defined by our group. The stars in NGC~6139 show a chemical
pattern normal for GCs, with a rather extended Na-O (and Mg-Al) anticorrelation.
NGC~6139 behaves like expected from its mass and contains a large fraction
(about two thirds) of second-generation stars.  
}

\maketitle

\begin{figure*}
\centering
\includegraphics[scale=0.9,bb=20 180 620 650,clip]{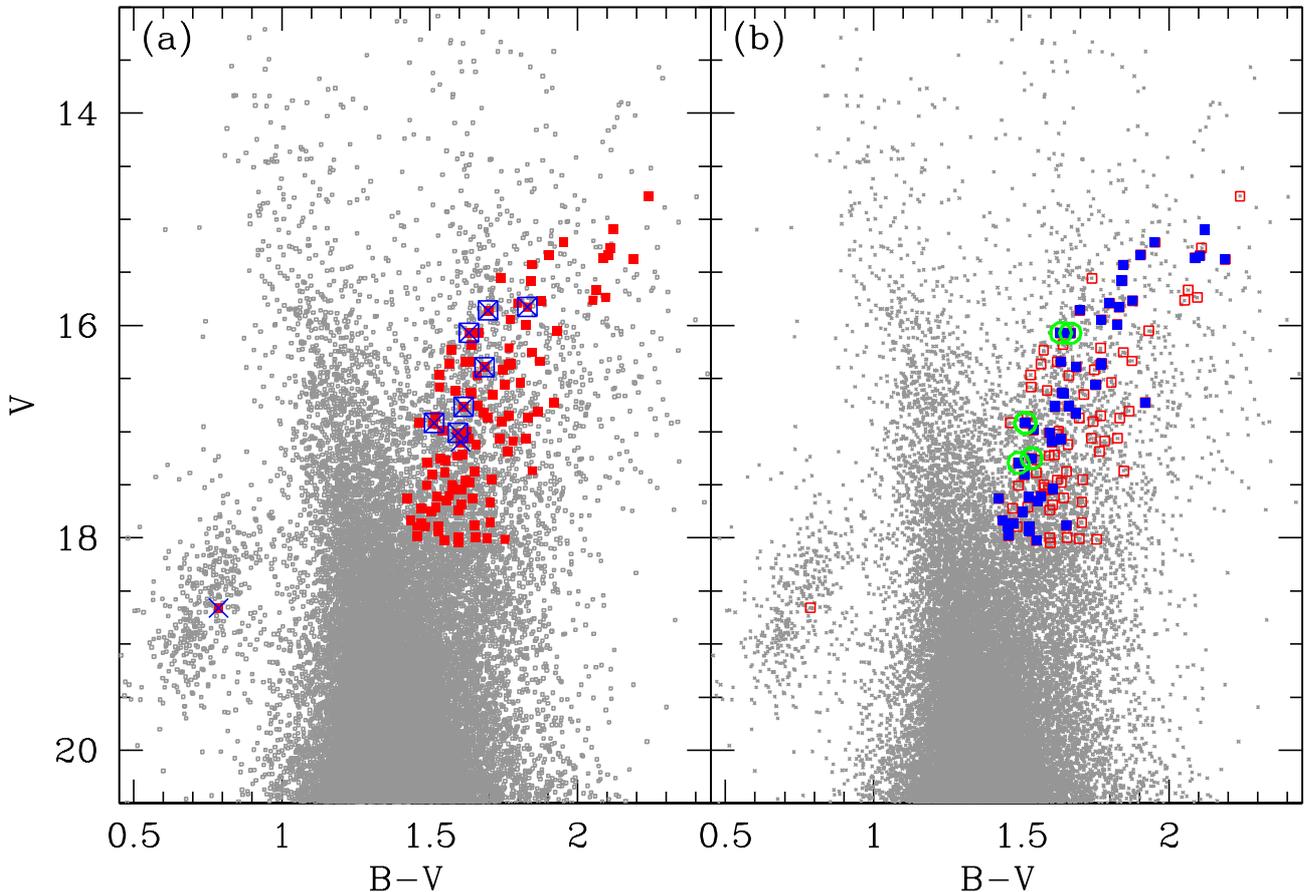} %cmdoss
\caption{(a) CMD of NGC~6139, with its quite short blue HB. 
In grey all stars, in red all targets
observed with FLAMES. The seven UVES stars are indicated by blue open squares; the stars with RV from \cite{saviane12} by blue crosses. (b) The non member stars (and the star on the HB) are indicated by red open squares, the stars candidate members on the basis of their RV by filled blue squares. The member stars define a reasonably tight RGB and we indicate with green circles the probable AGB stars (see Sec.~3.1)}
\label{cmdoss}
\end{figure*}

\section{Introduction}

Once considered as a good example of simple stellar populations, Galactic
globular clusters (GCs)  are currently thought to have formed in a complex chain
of events, which left a fossil record in their chemical composition  \citep[see
e.g., the review by][]{gratton12}.   Photometrically, GCs often exhibit  spread,
split and even multiple sequences that can be explained by different chemical
composition among cluster stars, in particular of light elements like He, C, N,
O  \citep[e.g.,][]{stromgren,sbordonestro,milone47t,piottouv}.  Our FLAMES 
survey of more than 20  Milky Way (MW) GCs \citep[see][and references
therein]{carretta09a,n4833} combined with literature data, demonstrated that
most, perhaps all, GCs host multiple stellar populations \citep[see][]{zcorri}.
Variations of Na and O (and sometimes of Mg and Al) abundances trace these
different sub-populations. 

Our large and homogeneous database allowed us a quantitative study of the Na-O 
anticorrelation. In all the analyzed GCs we found about one third of stars of
primordial composition,  similar to that  of field stars of similar metallicity
(low Na, high O), i.e., belonging to the first  generation (FG). The other two
thirds have a modified composition (increased Na, depleted O) and belong to the
second generation (SG) of stars, polluted by the FG \citep[see,
e.g.,][]{carretta09a,carretta09b}. Which were the more massive stars of FG that 
produced the gas of modified composition is still an unsettled question, see
e.g., \citet{decressin07,ventura01}, or \cite{bastian} for an alternative
view.  

We also found that the extension of the Na-O anticorrelation tends to be larger
for higher mass GCs  and that, apparently, there is an observed minimum cluster
mass for appearance of the  Na-O anticorrelation \citep{zcorri}.   It is
important to understand if this limit is  due to the small statistics (less
low-mass clusters have been studied,  and only a few stars were observed in
each). This in an important constraint for cluster formation mechanisms, because
it indicates the mass at which we expect that a cluster is able to retain part
of the ejecta of the FG, hence to show the Na-O signature (the masses of the
original clusters are expected to be much higher than the present ones, since
the SG has to be formed by the ejecta of the FG). 

Variations in Na, O, and He are tied, but do not tell us exactly the same 
story. The effects of increased  He are visible in colour-magnitude diagrams
(CMD) in the main sequence (MS) phase \citep[e.g.,][]{piotto07} or, more
evidently, in the horizontal branch phase (HB, e.g.,
\citealt{dantona05,grattonhb};  \citealt{milonehb}). Based on the possibility of reproducing their
HBs with a single He value, \cite{caloi11},  proposed that some GCs are composed
by a single generation (FG-only) or predominantly by a single generation
(mainly-FG). It would be important to measure their Na, O, and Al abundances to
clarify the issue. In fact Al, produced only at higher temperatures than Na 
\citep[e.g.,][for an application to the GC peculiar chemistry]{prantzos07}, i.e.
by more massive stars, should follow the He enrichment better than Na (and O).
We can then expect to find Na and O variations without significant He
enhancement; this would produce short HBs (and no variations in Al).\footnote This is
what seems to happen, for instance, for NGC~6397 and NGC~6838 \citep{carretta09b},
for which \cite{grattonhb} found no necessity of He dispersion to explain their HBs. Hovever, \cite{milone6397}, studying the main sequence of NGC~6397, found an internal variation in Y of about 0.01, small, but exceeding their measurement error. In
summary, we would be in presence of FG and SG stars even without significant He
variations; this is important to understand the cluster formation and early
intracluster gas pollution.

After studying the high-mass clusters, we begun a systematic study of low-mass
GCs, FG-only/mainly GCs, and high-mass and old open clusters (OCs). Our goal is
to empirically find the mass limit for the appearance of the Na-O
anticorrelation  and to ascertain whether there are differences between
high-mass and low-mass clusters' properties, e.g. in the relative fraction of FG
and SG stars. 

We also included GCs belonging to the Sagittarius  dwarf spheroidal (Sgr dSph)
to study if there are differences in GCs formed in different environments (the
MW and dwarf galaxies). In fact, GCs born in a dSph are expected to have
experienced  a milder tidal field thus possibly retaining a larger fraction of
their original mass.  Only a few old GCs in Fornax and LMC have their abundances
derived using high-res  spectroscopy. These GCs also seem to host two
populations  \citep{letarte06,johnsonLMC,mucciarelli09} but the fractions of FG
and SG stars in Fornax  and LMC GCs of similar mass seem different. Is this
again a problem of low statistics or is the galactic environment (a dwarf
spheroidal vs a dwarf irregular) influencing the GC formation mechanism? This is
outside the main theme of the present paper and we refer the interested reader
to the papers by \cite{larsen,dantonafnx} for a more detailed discussion.

We have already gathered high-resolution spectra of a low-mass GC (NGC~6535,
which will be analysed in a forthcoming paper), two high-mass OCs (Berkeley~39,
NGC~6791, see \citealt{be39,bragaglia6791}), a proposed FG-only GC (NGC~6139,
this paper),  and  Terzan~8 \citep{ter8},  belonging to the Sgr dSph.   In
Terzan~8 we see some indication of a SG, at variance with other low-mass Sgr GCs
(Ter~7,  Pal~12 \citealt{taut04,sbo07,cohen}) or to Rup~106 \citep{rup106}.  
However, in Terzan~8 the SG seems to be a minority component, contrary to what
occurs for  high-mass GCs. The Na-O anticorrelation had never been observed in
OCs  \citep{desilva,naoc} and we confirmed  this using large samples of
stars both for Berkeley~39 \citep{be39} using FLAMES spectra and  NGC~6791
\citep{bragaglia6791} using Hydra@WIYN and HIRES@Keck spectra.  In particular,
for NGC~6791, where \cite{geisler6791} claimed to have found some variations  in
Na, O, we did not detect any evidence of this trend. Our findings are
corroborated by independent works, see \cite{apo6791,hires6791}.

We concentrate here on the possibly FG-only cluster  NGC~6139.  In Section 2 we
present literature information on the cluster, in Section 3 we describe the
photometric data, the spectroscopic observations, and the derivation of
atmospheric parameters. The abundance analysis is presented in Section 4 and a
discussion on the light-element abundances is given in Section 5.

\section{NGC~6139 in literature} 

NGC~6139 is a massive GC (its absolute visual magnitude, a proxy for mass, is
$M_V=-8.36$, \citealt{harris}) located  toward the centre of the MW, at
$l=342.37\degr$, $b= 6.94\degr$. The cluster has received relatively scarce
attention in the past; in particular, it has never been studied with
high-resolution spectroscopy before.

\cite{hazen} found ten variable stars in NGC~6139, five of which are RR Lyrae
stars. The four RRab seem to indicate a Oosterhoff type II. A colour-magnitude
diagram (CMD) was presented by \cite{samus}; they employed photographic $B$ and
$V$ plates and their photometry reaches only the red giant branch (RGB) and HB.
They found the cluster quite metal-poor (close to  [Fe/H]=-2) from the slope of
the RGB, and highly reddened. \cite{zinnbarnes} used $VI$ photometry barely
reaching the main sequence turn-off, noted the high and differential reddening
(and corrected for a gradient), and determined [Fe/H]=$-1.71\pm 0.20$ and a mean
reddening of $E(V-I)=1.03\pm 0.04$, corresponding to $E(B-V)=0.76\pm0.03$.
Similar results were reached by \cite{ortolani} on the basis of $VI$ photometry
and \cite{davidge} who used near-infrared data and who also noted that the
differential reddening is not a significant problem, since the RGB sequence
becomes very well defined once the field stars contamination is statistically
removed. Finally, \cite{snapshot} presented WFPC2 photometry, as part of their
HST snapshot programme (74 MW GCs observed with the $F439W$ and $F555W$ bands)
the RGB and HB are very well defined, and the main sequence turn-off is better
defined than in previous works. Their data are publicly available and we used
them for the present work (see next Section)

The only determination of metallicity based on spectroscopy is by
\cite{saviane12}, who used  spectra at resolution $R\sim2500$ in the region of
the infrared Ca {\sc ii} triplet (CaT). NGC~6139 is one of the 20 GCs in the
paper. They observed 19 stars, 15 of which were considered members (we used this
information to select our targets, see next Section).  The metallicity they
obtained from the CaT lines is [Fe/H]=$-1.63$, rms=0.13 (on the metallicity
scale defined in \citealt{carretta09c}).

\cite{caloi11} include NGC~6139 among their list of candidate FG clusters. They
tried to identify those clusters whose HB could be reproduced by a single mass
(in the framework where the dispersion in mass of HB stars is due -at least in
part- to variations in He content, in turn indication of multiple generations).
A first clue is the short HB  (similar to the HB of NGC~6397, see \citealt{snapshot}, which, however has both a ``normal" Na-O anticorrelation,
and a small $\Delta$Y, see \citealt{carretta09a,carretta09b,milone6397}, respectively). However, given the high mass ($\log
M=5.58~M_\odot$, their Table~1), NGC~6139 should have produced a SG, but failed
to lose a large fraction of the FG stars, so it should be presently
FG-dominated. \cite{caloi11}  do not discuss in detail the case for NGC~6139,
but its inclusion in the list, combined with the absence of previous
high-resolution spectroscopic studies made it a good target for our on-going
programme.

\begin{table}
\centering
\setlength{\tabcolsep}{1.3mm}
\caption{Log of FLAMES observations.}
\begin{tabular}{lccccc}
\hline
Setup   &  UT Date   &  UT$_{init}$ & exptime & airmass & seeing\\
        & (yyyy-mm-dd) & (hh:mm:ss) & (s)	   & 	    & (arcsec) \\
\hline
HR11  & 2014-05-10 & 02:25:58.027 & 3600  & 1.459 &  0.76\\
HR11  & 2014-05-10 & 03:27:50.586 & 3600  & 1.217 &  0.88\\
HR11  & 2014-05-10 & 04:29:58.656 & 3600  & 1.091 &  0.64\\
HR11  & 2014-05-10 & 05:36:41.844 & 3600  & 1.035 &  0.68\\
HR11  & 2014-05-10 & 06:50:54.214 & 3600  & 1.052 &  0.57\\
HR11  & 2014-05-10 & 07:54:38.745 & 3600  & 1.136 &  0.47\\ 
HR13  & 2014-07-22 & 03:21:00.913 & 3600  & 1.164 &  1.06\\
HR13  & 2014-07-22 & 04:24:36.587 & 3600  & 1.366 &  1.09\\
HR13  & 2014-07-25 & 03:14:57.248 & 3600  & 1.178 &  0.76\\
HR13  & 2014-07-26 & 04:12:36.160 & 3600  & 1.383 &  0.59\\
HR21  & 2014-06-03 & 01:01:28.004 & 3090  & 1.409 &  0.86\\
HR21  & 2014-07-03 & 00:29:34.582 & 3090  & 1.137 &  0.99\\
\hline
\end{tabular}
\label{log}
\end{table}

\section{Observations and analysis} 

Of the photometric data discussed in previous section, only the HST catalogue is
publicly available. However, the very small field of view (FoV, about 2\arcmin \
side) is not well suited to the FLAMES FoV (25\arcmin \ diameter). We then
retrieved from the ESO archive some $B$ and $V$ filter frames acquired (under
program 68.D-0265) with the Wide Field Imager (WFI) at the 2.2m ESO-MPG
telescope, which has a FoV of about 30\arcmin \ side.

They were reduced in the standard way, correcting for bias and flat field using
IRAF.\footnote{IRAF is distributed by the National Optical Astronomical
Observatory, which are operated by  the Association of Universities for 
Research in Astronomy, under contract with the  National Science  Foundation.} 
Stars were detected independently on the $B$ and $V$ frames and the instrumental
magnitudes were obtained using the Point Spread Function (PSF) fitting code
DAOPHOT-II/ALLSTAR \citep{ste1,ste2}.   We employed the \emph{2 Micron All Sky
Survey}  Catalogue (2MASS, \citealt{2mass})   and the CataXcorr
code\footnote{http://www.bo.astro.it/$\sim$paolo/Main/CataPack.html},  developed
by P. Montegriffo, to compute the astrometric solution and transform the
instrumental pixel coordinates into J2000 celestial coordinates. The astrometric
precision is about 0.2 arcsec, perfectly compatible with the requirements of the
FLAMES observations. No standard stars were available, so we calibrated our
photometry to the HST one using the stars in common. The final photometric
catalogue will be made available through the CDS.

The resulting CMD is shown in Fig.~\ref{cmdoss}. As expected from its Galactic
position, the  field stars contamination is conspicuous, but the cluster RGB and
HB are visible, especially when restricting to the very central region. The
cluster sequences are also affected by differential reddening (DR), as already
discussed in literature. This  could be relevant for the spectroscopic analysis,
since our atmospheric parameters are derived from photometry.  However, as
already done for other difficult cases, resorting to optical-IR colours (in
particular, $V-K$), greatly alleviates the problem. For instance, the bulge GC
GC~6441 \citep{gra6441} has an rms scatter of 0.05 mag in E($B-V$), resulting 
in a random (star-to-star) uncertainties in the effective  temperatures of
$\pm80$~K. We roughly evaluated the size of DR for NGC~6139 by defining a red
giant branch (RGB) ridge line with the help of the HST photometry and projecting
the candidate members (see next section) on it along the reddening vector. The
average  displacement required is 0.03 mag so we expect a possible error of the
order of 50 K.

\begin{figure}
\centering
\includegraphics[scale=0.8,bb=40 150 400 720,clip]{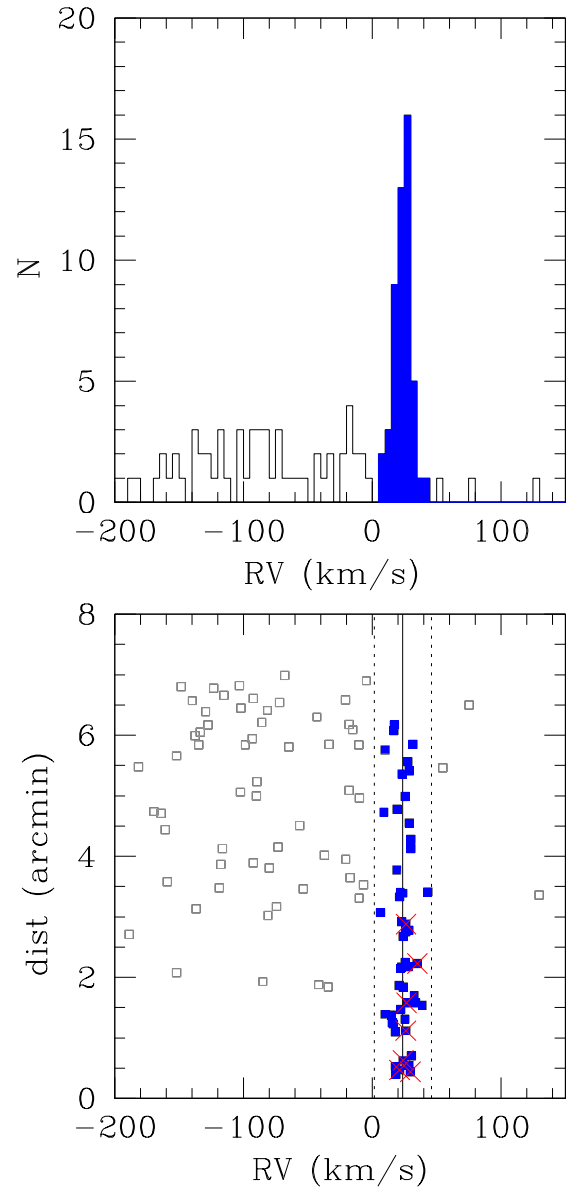} %plotrv.ps
\caption{Upper panel: histogram of all measured RVs (open histogram) and of the
candidate member stars only (blue filled histogram). Lower panel: RVs versus
distance from the cluster centre. The candidate member stars are indicated by
filled blue squares, the UVES targets by large red crosses, while the lines
indicate the cluster mean RV and the $\pm 3\sigma$ interval.}
\label{plotrv}
\end{figure}

\subsection{FLAMES spectra}
NGC~6139 was observed with the multi-object  spectrograph FLAMES@VLT
\citep{pasquini}. We used the GIRAFFE high-resolution setups HR11, HR13, and
HR21  (R=24200, 22500, 16200, respectively) which contain two Na doublets, the
[O {\sc i}]  line at 6300 \AA, and the Al doublet at 8772 \AA, plus several Mg
lines. The observations were performed in service mode; a log is presented in
Table~\ref{log}. The HR11 and HR13 GIRAFFE observations were coupled with the
high-resolution (R=47000) UVES 580nm setup ($\lambda\lambda\simeq4800-6800$~\AA)
and the HR21 observations with the UVES 860nm setup
($\lambda\lambda\simeq6600-10600$~\AA).  We will use only the 580nm spectra in
this paper, for uniformity with our previous works and because they have better
signal to noise.

The $B,V$ WFI catalogue was used to select candidate targets for the
spectroscopic observations. As done for all other GCs in our sample, we chose
only stars without close neighbours ($\le2.5\arcsec$). We cross-identified the
stars in \cite{saviane12} to allocate seven UVES fibres to high probability
cluster members; the eigth fibre was put on an empty spot for sky subtraction. 
We allocated 103 GIRAFFE fibres to RGB stars, one to a HB star, and 16 to sky
positions. Virtually all targets are brighter than the RGB bump
($V=17.867\pm0.019$; \citealt{nataf}). The 111 stars observed are shown in 
Fig.~\ref{cmdoss}(a) with different symbols; their coordinates and magnitudes
are given in Table~\ref{tabM} (for candidate members, based on their RV) and
Table~\ref{tabNM} (for non members). All stars observed are within 7 arcmin form
the centre, see Fig.~\ref{plotrv}, lower panel, i.e., well within  the tidal
radius (10.5 armin, \citealt{harris}). Almost all stars observed are beyond the
half-mass radius (0.85 arcmin, Harris).

The spectra were reduced using the ESO pipelines for UVES-FIBRE and GIRAFFE
data; they take care of bias and flat field correction, order tracing,
extraction, fibre transmission, scattered light, and wavelength calibration. We
then used IRAF routines on the 1-d, wavelength-calibrated individual spectra to
subtract the (average) sky, shift to zero RV, and combine all the exposures for
each star. The region near the [O {\sc i}] line  required special attention because of the strong sky emission and the many absorptions. Given the low RV of the cluster, the HR13 exposures were scheduled when the Earth motion took the sky emission farther away from the stars' [O {\sc i}] line. We further checked that in all these exposures the telluric absorptions did not fall on the stars' line, by visual comparison with a telluric template.
  The RV was measured using DOOp
\citep{doop}, an automated wrapper for DAOSPEC \citep{daospec}; the average,
heliocentric value for each star is given in Tables~\ref{tabM} and \ref{tabNM}. 

We show in Fig.~\ref{plotrv} the histogram of the RVs and a plot of RVs versus
distance from cluster centre (taken by \citealt{harris}, 2010 web update); the
cluster signature is clear. Our is the first estimate of RV based on a large
numebr of stars observed at high-resolution; we found $\langle RV\rangle
=+28.88$, rms=7.34 km~s$^{-1}$ (a typical rms, for a GC). This value is
compatible with the RV=6.7$\pm$6.0 km~s$^{-1}$ in \cite{harris}, based on much
lower resolution and quality data\footnote{ \cite{harris} refers to
\cite{webbink}, who uses RVs from \cite{kinman}, derived on spectra at
$R\sim200$ of four stars.}, and is in good agreement with \cite{saviane12}, who
give RV=34$\pm4$ km~s$^{-1}$, based on 15 stars (for a direct comparison of 
RVs  for eight stars in common, see Table~\ref{tabM}). 

We are left with 50 candidate members based on the RV, i.e., falling within
$\pm3\sigma$ of the cluster average. After pruning the sample, the RGB of
NGC~6139 is well defined and tight, even in presence of DR (see
Fig.~\ref{cmdoss}(b)). In the following we will discuss only them.

Following our well tested procedure \citep[for a lengthy description, see
e.g.,][]{carretta09a,carretta09b}, effective temperatures $T_{\rm eff}$\ for our
targets were derived using an average relation between apparent magnitudes and
first-pass temperatures from $V-K$ colours and the calibrations of
\cite{alonso,alonso2}. This method permits to decrease the star-to-star errors
in abundances due to uncertainties in temperatures, since magnitudes are less
affected by  uncertainties than colours. This is particularly true for NGC~6139,
which presents high and variable reddening, for which we used the apparent $K$
magnitudes in our relation with  $T_{\rm eff}$ because the impact of the DR on
these magnitudes is smaller.   The adopted reddening $E(B-V)=0.75$ and distance
modulus $(m-M)_V=17.35$ are from the \cite{harris} catalogue. Gravities were
obtained from apparent magnitudes and distance modulus, assuming the  bolometric
corrections from \cite{alonso}. We adopted a mass of 0.85~M$_\odot$\  for all
stars and $M_{\rm bol,\odot} = 4.75$ as the bolometric magnitude for the Sun, as
in our previous studies.

We eliminated trends in the relation between abundances from Fe~{\sc i} lines
and expected  line strength \citep{magain} to obtain values of the
microturbulent velocity $v_t$.

Finally, using the above values we interpolated  within the \cite{kur} grid of
model atmospheres (with the option for overshooting on) to derive the final
abundances, adopting for each star the model with the appropriate atmospheric
parameters and whose abundances matched those derived from  Fe {\sc i} lines.
Five stars among the candidate members turned out to have a metallicity higher
by more than 0.5 dex than the average of the others; they are probably field
stars and have been excluded from the most secure sample of RV $and$ metallicity
members. The atmospheric parameters for all bona fide member stars are given in
Table~\ref{tabatmo}.

Six of the candidate members (two with UVES and four with GIRAFFE spectra)
turned out too metal-poor (by about 0.15 dex); their position
in the CMD seemed to indicate that they could be Asymptotic Giant Branch stars
(AGB) and not RGB. Following the same procedure used for the RGB stars, we
derived a separate colour-temperature relation for them, finding temperatures
higher by about 110~K. Adopting the new temperatures we repeated the analysis,
finding a metallicity in very good agreement with the cluster average. These
second  values for $T_{\rm eff}$ and metallicity are given in
Table~\ref{tabatmo}.

\begin{figure}
\centering
\includegraphics[scale=0.75, bb=10 150 400 710,clip]{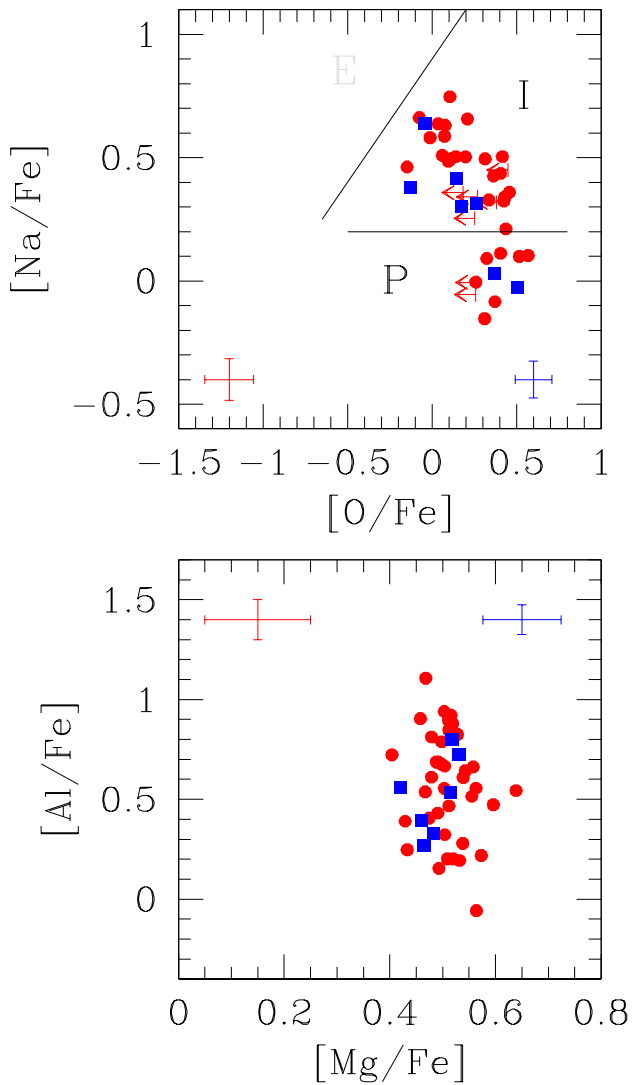} %anticorr
\caption{Anticorrelations between light elements: Na-O (upper panel) and Mg-Al
(lower panel). GIRAFFE stars are indicated by red filled circles and UVES stars
by filled blue squares; upper limits in O are shown as arrows. The errorbars are
the cluster average internal (star-to-star) errors for each element for the
GIRAFFE sample  in red and UVES in blue (see text). In the upper panel we indicate the separation
between P and I, and I and E populations, according to \cite{carretta09a}.}
\label{anti}
\end{figure}

\section{Abundances}

Beside Fe, we present here abundance of O, Na, Mg, and Al. The last was derived
from the Al~{\sc i} doublet at 6696-98 \AA \ for stars observed with UVES, and
from the doublet at 8772 \AA \ for stars observed with GIRAFFE. The abundance
ratios for all elements are given in Table~\ref{tablight}, together with number
of lines used and rms scatter.

The abundances were derived  using equivalent widths (EW).  We measured the EW
of iron and other elements using the code ROSA \citep{rosa} adopting a
relationship between EW and FWHM (for details, see \citealt{bragaglia01}). The
atomic data for all the lines in the UVES spectra and in the HR11, HR13 setups
and the solar reference values  come from \cite{gratton03}.  The Na abundances
were corrected for departure form local thermodynamical equilibrium according to
\cite{gratton99}.

 The derived Fe abundances do not show any trend with $T_{\rm eff}$. The average
metallicity derived from stars with UVES spectra  is  [Fe/H]$=-1.579 \pm 0.015
\pm 0.058$ (rms = 0.040 dex, 7 stars) from neutral species, where the first
error is from statistics and the second is systematic (see below for the
computation). For stars with GIRAFFE spectra we found a very similar value:
[Fe/H]$=-1.596 \pm 0.006 \pm 0.042$ (rms = 0.038 dex, 38 stars). The abundance
derived from single ionized species is in very good agreement. We found
[Fe/H]{\sc ii}$=-1.541 \pm 0.016 \pm 0.054$ (rms = 0.043 dex, 7 stars) and
[Fe/H]{\sc ii}$=-1.579 \pm 0.008 \pm 0.042$ (rms = 0.048 dex, 38 stars) for UVES
and GIRAFFE, respectively. This supports the adopted temperature scale and
gravities.

Ours is the first high-resolution spectroscopic study of this cluster, so no
real comparison with previous determinations is possible. However, the
metallicity we find is in good agreement with the one based on CaT
([Fe/H]$=-1.63$, \citealt{saviane12}) and with the general finding of a low
metallicity ($\sim -1.7$ or $-2$) based on photometry.

To estimate the error budget we closely followed the procedure described in
\cite{carretta09a,carretta09b}. Table~\ref{t:sensitivityu39} (for UVES spectra)
and Table~\ref{t:sensitivitym39} (for GIRAFFE spectra) provide the sensitivities
of abundance ratios to errors in atmospheric parameters and EWs and the internal
and systematic errors.   For systematic errors we mean 
the errors that are different for the various GCs considered in our series and
which produce scatter in relations involving different  GCs; however, they do
not affect  the star-to-star scatter in NGC~6139. The cluster uncertainty in
T$_{\rm eff}$ can be estimated by multiplying the slope of the relation T$_{\rm
eff} - (V-K)$ in NGC~6139 and the  uncertainty in $E(V-K)$ (assumed to be 0.055
mag). We also quadratically sum a  contribution from a conservative estimate of
0.02 mag error in the zero point of $V-K$ colour. The resultant uncertainty is
propagated together with those in  distance modulus and stellar mass to estimate
the systematic uncertainty in surface gravity, while the systematic error in
$v_t$ is simply obtained by  dividing the internal error for the square root of
the number of stars. The cluster (systematic) error in the metallicity for
NGC~6139 is then obtained by the quadratic sum of the above three terms
(multiplied for the proper  sensitivity)  with the statistical errors of
individual abundance determination. The sensitivities were obtained by
repeating the abundance analysis for all stars,  while changing one atmospheric
parameter at the time, then taking the average; this was done separately for
UVES and GIRAFFE spectra. The amount of change in the input parameters used in
the sensitivity computations is given in the Table header.

\begin{table*}
\setcounter{table}{5}
\centering
\caption[]{Sensitivities of abundance ratios to variations in the atmospheric
parameters and to errors in the equivalent widths, and errors in abundances for
stars of NGC~6139 observed with UVES.}
\begin{tabular}{lrrrrrrrr}
\hline
Element     & Average  & T$_{\rm eff}$ & $\log g$ & [A/H]   & $v_t$    & EWs     & Total   & Total      \\
            & n. lines &      (K)      &  (dex)   & (dex)   &kms$^{-1}$& (dex)   &Internal & Systematic \\
\hline        
Variation&             &  50           &   0.20   &  0.10   &  0.10    &         &         &            \\
Internal &             &   4           &   0.04   &  0.04   &  0.06    & 0.01    &         &            \\
Systematic&            &  46           &   0.06   &  0.06   &  0.02    &         &         &            \\
\hline
$[$Fe/H$]${\sc  i}& 86 &    +0.060     & $-$0.002 &$-$0.008 & $-$0.028 & 0.011  &0.021    &0.058	\\
$[$Fe/H$]${\sc ii}& 16 &  $-$0.031     &   +0.086 &  +0.022 & $-$0.014 & 0.026  &0.034    &0.042	\\
$[$O/Fe$]${\sc  i}&  1 &  $-$0.044     &   +0.084 &  +0.038 &	+0.027 & 0.104  &0.108    &0.096	\\
$[$Na/Fe$]${\sc i}&  2 &  $-$0.016     & $-$0.038 &$-$0.015 &	+0.022 & 0.074  &0.075    &0.088	\\
$[$Mg/Fe$]${\sc i}&  2 &  $-$0.016     & $-$0.009 &$-$0.000 &	+0.012 & 0.074  &0.074    &0.021	\\
$[$Al/Fe$]${\sc i}&  2 &  $-$0.017     & $-$0.008 &$-$0.002 &	+0.026 & 0.074  &0.075    &0.077	\\
\hline
\end{tabular}
\label{t:sensitivityu39}
\end{table*}

\begin{table*}
\setcounter{table}{6}
\centering
\caption[]{Sensitivities of abundance ratios to variations in the atmospheric
parameters and to errors in the equivalent widths, and errors in abundances for
stars of NGC~6139 observed with GIRAFFE.}
\begin{tabular}{lrrrrrrrr}
\hline
Element     & Average  & T$_{\rm eff}$ & $\log g$ & [A/H]   & $v_t$    & EWs     & Total   & Total      \\
            & n. lines &      (K)      &  (dex)   & (dex)   &kms$^{-1}$& (dex)   &Internal & Systematic \\
\hline        
Variation&             &  50           &   0.20   &  0.10   &  0.10    &         &         &            \\
Internal &             &   4           &   0.04   &  0.04   &  0.13    & 0.02    &         &            \\
Systematic&            &  46           &   0.06   &  0.05   &  0.02    &         &         &            \\
\hline
$[$Fe/H$]${\sc  i}& 35 &    +0.058     & $-$0.001 &$-$0.007 & $-$0.025 & 0.022  &0.041    &0.054	\\
$[$Fe/H$]${\sc ii}&  2 &  $-$0.035     &   +0.088 &  +0.020 & $-$0.007 & 0.067  &0.101    &0.042	\\
$[$O/Fe$]${\sc  i}&  1 &  $-$0.041     &   +0.081 &  +0.034 &	+0.023 & 0.095  &0.144    &0.053	\\
$[$Na/Fe$]${\sc i}&  3 &  $-$0.016     & $-$0.040 &$-$0.018 &	+0.017 & 0.067  &0.084    &0.043	\\
$[$Mg/Fe$]${\sc i}&  2 &  $-$0.021     & $-$0.009 &$-$0.003 &	+0.013 & 0.095  &0.100    &0.021	\\
$[$Al/Fe$]${\sc i}&  2 &  $-$0.030     & $-$0.006 &$-$0.002 &	+0.019 & 0.095  &0.101    &0.051	\\
\hline
\end{tabular}
\label{t:sensitivitym39}
\end{table*}

\begin{figure*}
\centering
\includegraphics[scale=0.8, bb=40 100 680 450,clip]{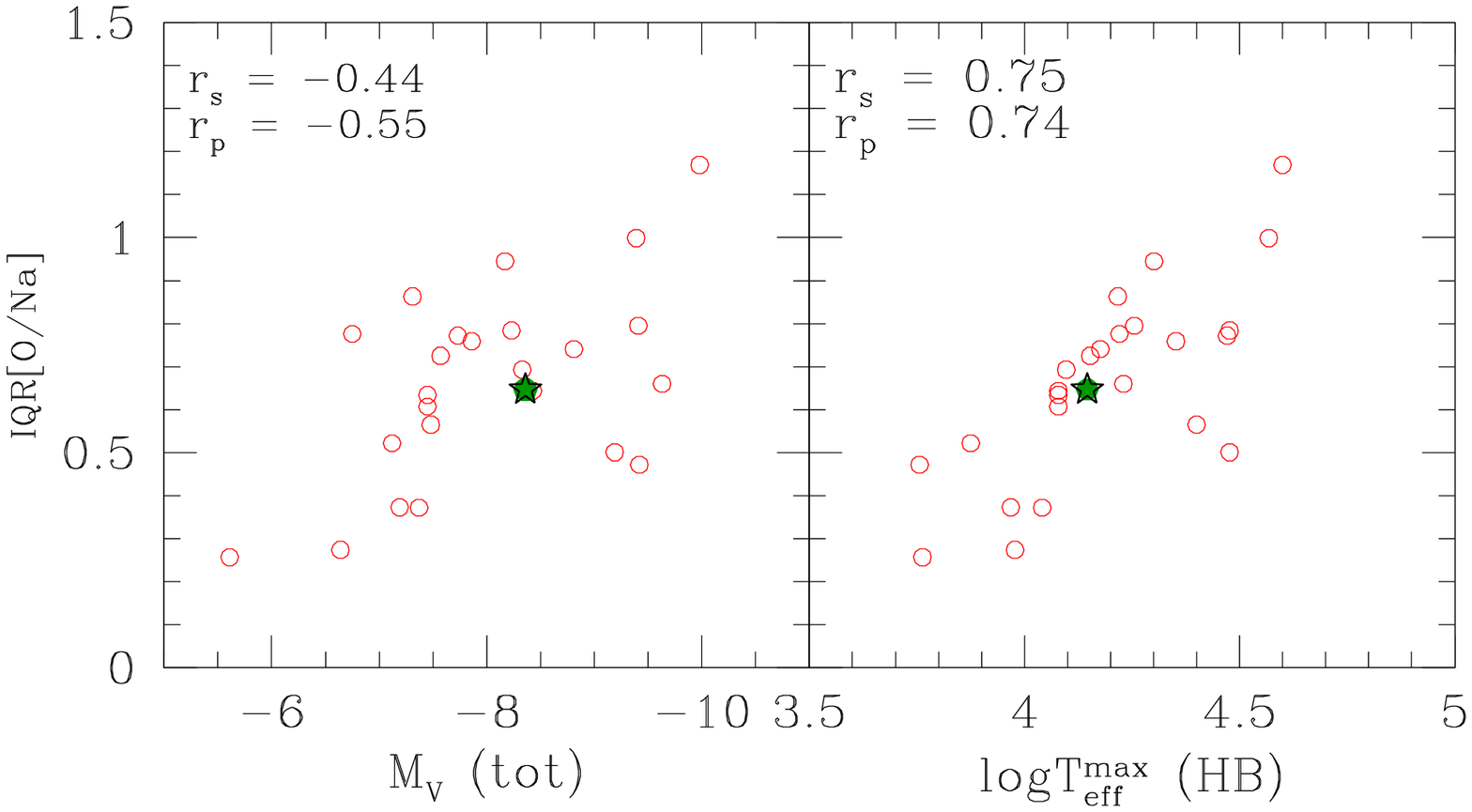} %iqr_mv_teff
\caption{Left-hand  panel: correlation between total absolute magnitude and the
interquartile range of the [O/Na] ratio for GC in our FLAMES survey \citep[][and
references therein]{carretta09a,m54,n1851,n362,n4833,m80}. Right-hand panel:
relation between the maximum temperature along the HB
\citep{recioblanco,carretta07} and IQR[O/Na]. In each panel NGC~6139 indicated
by a filled green star.  The Spearman rank correlation coefficient $r_s$ and the
Pearson linear correlation coefficient $r_p$ are listed in each panel.}
\label{iqr}
\end{figure*}

\section{Light elements anti-correlations}

Our final sample consists of 45 giant stars (seven observed with UVES and 38
with GIRAFFE, no stars are in common). We could measure Na, Mg, and Al in all
spectra (UVES and GIRAFFE), while O was measured in all seven UVES stars but
only in 36 GIRAFFE stars (29 actual detection and seven upper limits).

The resulting  relations between O and Na, Mg and Al are shown in
Fig.~\ref{anti}, where we clearly see the normal pattern of light elements
abundance ratios in GCs. NGC~6139 displays a Na-O anticorrelation like all
massive GCs studied to date. Interestingly, it also shows a
large variation in Al coupled with a moderate variation in Mg,
A variation in Al is not seen in all GCs \citep[see Fig.~6 in][]{carretta09b};  however, also when Al varies 
the moderate variation in Mg is typical, with only rare exceptions such as e.g., NGC~2808, NGC~6752
\citep{carretta2808,carretta6752}.
On the other hand, we observe a clear Na-Al correlation, with a Pearson 
correlation coefficient $r_P=0.70$, which is very 
significant.
This  confirms that the Na-Na and Mg-Al cycles are related, 
but also suggests that they do not occur exactly in the same polluting stars.

Using the criteria defined in \cite{carretta09a} we can
define the fraction of stars belonging to the primordial (P) and
second-generation (I, E) components. The fraction of first-generation P stars
is   $26\% \pm 8\%$; the fraction of second-generation I stars (with moderate
modification of the abundances) is $74\% \pm13\%$; finally, there are no
second-generation E stars (with extremely modified composition) in our sample
of 43 stars in NGC~6139. These numbers agree with what is found in all massive
Milky Way GCs \citep[see e.g.,][and references therein]{zcorri,n4833,m13}; SG
stars are presently the dominant population in them. Our study then does not
confirm NGC~6139 as a  mainly-FG cluster, as proposed by \cite{caloi11}.

As it happened for the other GCs in our sample \citep[e.g.,][]{carretta09a},
where most of the abundance ratios are based on GIRAFFE spectra, the errors on
abundances produce almost continuous distributions (Fig.~\ref{anti}). However, a
separation between the P and I stars is visible in the upper panel (Na and O),
especially for the stars observed with UVES. A clearcut separation in groups
along the Na-O anticorrelation has been seen only in high-quality and resolution
data, such as the UVES spectra in M4 \citep{marinom4} and NGC~2808
\citep{carretta2808}.  The Mg-Al distribution (Fig.~\ref{anti} lower panel)
looks continuous, with a total excursion of about 1 dex in [Al/Fe].

The Na-O anticorrelation has  moderate extension, with an interquartile range
for the [O/Na] ratio \citep{carretta06} IQR[O/Na]=0.647. This IQR makes NGC~6139
fit in the relations with absolute magnitude ($M_V$, i.e., mass) and maximum
temperature reached on the HB ($T_{eff}^{max}$(HB),
\citealt{recioblanco,carretta07}, see Fig.~\ref{iqr} for a graphical
representation). 

In conclusion, with our study of 45 member stars in NGC~6139 conducted with
FLAMES spectra we determined for the first time its metallicity  with
high-resolution spectroscopy (the cluster has an intermediate metallicity, with
[Fe/H]=--1.58 dex). We also measured the light elements O, Na, Mg, and Al
involved in proton-capture reactions of H-burning at high temperatures; we
detected the usual correlations and anticorrelations found in almost all GCs
investigated so far. The ratio of first-to-second generation stars in NGC~6139
is typical of the majority of GCs and is consistent with its relatively high
total present-day mass. We do not support the idea that NGC~6139 is a
FG-dominated cluster. The intermediate extension of the Na-O anticorrelation, as
measured by the interquartile range of the [O/Na] ratio, fits very well the
relation with the maximum temperature along the HB found by \cite{carretta07}. 
All in all NGC~6139 behaves like a normal MW GC of similar mass and HB
extension. The formation event that shaped its internal chemistry  seems
consistent with the pattern of a regular, relatively high-mass, blue HB GC.

 However, the finding of a large variation in Al, which should also imply
some spread in He, does not perfectly fit with the HB of the cluster.  We
calculated IQR[Al/Mg], finding a value of 0.425; according to Fig.~28 in
\cite{grattonhb}, this would require a $\Delta$Y of about 0.02 (but note that
the relation between IQR[Al/Mg] and $\Delta$Y is rather poorly determined, being
based on only eight points). We note that also the more massive, metal-poorer
NGC~5024 (M53, with $M_V=-8.71$, [Fe/H]=-2.10, \citealt{harris}), proposed as a
FG-dominated GC by \cite{caloi11} on the basis of its HB and discussed at length
in that paper, has been demonstrated to host a substantial SG by
\cite{apogee15}, using APOGEE data. NGC~5024 has a large spread in Al (with a
bimodal  distribution) and almost no Mg variation; it behaves like NGC~5272
(M3,  $M_V=-8.88$, [Fe/H]=-1.50, \citealt{harris}), another GC with  a rather
short HB, even more massive than NGC~6139, but with a similar metallicity. The
relation between Al and He is evidently not completely understood. As  as an
additionally cautionary note, \cite{bastian15} claim that no one of the 
enrichment mechanisms proposed to explain multiple populations in GCs can
consistently do so, because all predict too large He spreads
in order to reproduce the observed spreads in light elements, notably 
those in Na and O. More observational constraints are required, both from
spectroscopy with large samples of stars for which elements of all
nucleosynthetic chains are measured (like we and other groups are doing for an
increasing number of of GCs) and from precise photometry \citep[e.g., ][to
determine He]{piottouv}. These constraints will be helpful to improve modelling
the formation and early enrichment phases of GCs.

\begin{acknowledgements}
We thank the referee for constructive comments which improved the paper.
This research has made use of WEBDA, SIMBAD database, operated at CDS,
Strasbourg, France, and NASA's Astrophysical Data System. This publication makes
use of data products from the Two Micron All Sky Survey, which is a joint
project of the University of Massachusetts and the Infrared Processing and
Analysis Center/California Institute of Technology, funded by the National
Aeronautics and Space Administration and the National Science Foundation.  This
research has been partially funded by PRIN INAF 2011 "Multiple populations in
globular clusters: their role in the Galaxy assembly", and PRIN MIUR 2010-2011,
project ``The Chemical and Dynamical Evolution of the Milky Way and Local Group
Galaxies''. 
\end{acknowledgements}

\clearpage

\begin{table*}
\centering 
\setcounter{table}{1}
\caption{Information on the member stars observed.
{\bf the complete table is available in the on-line version}}
\begin{tabular}{ccccccrcl}
\hline
\hline
 ID   &    RA       &  Dec         &  V     &  B      &  K	  &  RV    & err  & Notes\\
      & (hh:mm:ss)  & (dd:pp:ss)   &        &         & (2MASS)& (km~s$^{-1}$) & (km~s$^{-1}$)& \\
\hline
\multicolumn{9}{c}{UVES} \\
t00540 & 16:27:39.82 & -38:50:18.05  & 15.825 & 17.656 & 10.601  &   24.00 &  0.30 &   4, RV$_{S12}$=37 \\
t00553 & 16:27:39.22 & -38:51:18.04  & 15.859 & 17.557 & 10.856  &   29.63 &  0.60 &   1, RV$_{S12}$=46 \\
t00670 & 16:27:42.77 & -38:50:56.29  & 16.069 & 17.703 & 11.329  &   21.27 &  0.50 &   3, RV$_{S12}$=36, AGB \\
t00951 & 16:27:35.86 & -38:53:40.28  & 16.391 & 18.077 & 11.706  &   26.21 &  0.60 &  13, RV$_{S12}$=44 \\
t01348 & 16:27:40.82 & -38:48:42.10  & 16.764 & 18.379 & 12.049  &   35.00 &  0.30 &  11, RV$_{S12}$=19 \\
t01554 & 16:27:38.41 & -38:49:23.44  & 16.920 & 18.434 & 12.261  &   26.71 &  1.60 &   8, RV$_{S12}$=38, AGB \\
t01699 & 16:27:42.55 & -38:49:53.35  & 17.010 & 18.605 & 12.383  &   26.09 &  0.50 &   7, RV$_{S12}$=42 \\
\multicolumn{9}{c}{GIRAFFE, members} \\
t00261 & 16:27:43.32 & -38:57:04.20  & 15.097 & 17.217 &  9.530  &   17.37 &  0.35 &	       \\
t00284 & 16:27:46.06 & -38:54:03.74  & 15.218 & 17.170 &  9.988  &   21.40 &  0.55 &	       \\
t00316 & 16:27:46.75 & -38:51:04.61  & 15.337 & 17.240 & 10.145  &   15.68 &  0.34 &	       \\
t00329 & 16:27:20.70 & -38:53:22.86  & 15.341 & 17.445 &  9.744  &   28.93 &  0.42 &	       \\
t00337 & 16:27:30.42 & -38:52:04.70  & 15.363 & 17.450 &  9.754  &   25.63 &  0.84 &	       \\
\hline
\end{tabular}
\begin{list}{}{}
\item[] In Notes we give the number and RV in Saviane et al. (2012);
\item[] Star t06169 is on the HB.
\end{list}
\label{tabM}
\end{table*}

\begin{table*}
\centering 
\setcounter{table}{2}
\caption{Information on non member stars on the basis of RV or metallicity, all observed with GIRAFFE.
{\bf the complete table is available in the on-line version}}
\footnotesize
\begin{tabular}{ccccccrcl}
\hline
\hline
 ID   &    RA       &  Dec         &  V     &  B      &  K	  &  RV    & err  & Note \\
      & (hh:mm:ss)  & (dd:pp:ss)   &        &         & (2MASS)& (km~s$^{-1}$) & (km~s$^{-1}$) &\\
\hline
t00187 & 16:27:25.45 & -38:48:42.66  & 14.781 & 17.020 &  7.078  &  -16.98 &  1.00 & 	RV  \\
t00306 & 16:27:10.97 & -38:53:19.81  & 15.271 & 17.381 &  9.388  &  -85.72 &  0.49 & 	RV  \\
t00392 & 16:27:48.91 & -38:51:43.38  & 15.555 & 17.294 & 10.639  &  -34.11 &  0.99 & 	RV  \\
t00466 & 16:28:07.92 & -38:49:47.99  & 15.666 & 17.730 &  9.262  & -181.62 &  0.69 & 	RV  \\
t00513 & 16:28:00.26 & -38:52:21.00  & 15.737 & 17.832 &  9.739  & -116.30 &  0.48 & 	RV  \\
t00526 & 16:27:45.67 & -38:55:14.59  & 15.764 & 17.815 &  9.802  & -160.69 &  0.52 & 	RV  \\
\hline
\end{tabular}
\label{tabNM}
\end{table*}

\setcounter{table}{3}
\begin{table*}
\centering
\caption[]{Adopted atmospheric parameters and derived metallicity for confirmed member stars.{\bf the complete table is available in the on-line version}}
\begin{tabular}{rccccrcccrccc}
\hline
Star   &   $T_{\rm eff}$ &  $\log$ $g$ &  [A/H]  & $v_t$	     &  nr &  [Fe/H]{\sc i} &  $rms$ &  nr &  [Fe/H{\sc ii} &  $rms$  \\
       &      (K)	&   (dex)     &  (dex)  & (km s$^{-1}$) &     &  (dex) 	  & 	  &     &  (dex)               \\
\hline
\multicolumn{11}{c}{UVES} \\
t00540 & 4156 &0.95 &-1.55 &1.85 &121 &-1.545 &0.089 &20 &-1.551 &0.076 \\ 
t00553 & 4212 &1.07 &-1.61 &1.40 &107 &-1.610 &0.102 &20 &-1.505 &0.131 \\ 
t00670 & 4417 &1.28 &-1.57 &1.43 & 91 &-1.570 &0.112 &13 &-1.566 &0.164 \\ 
t00951 & 4398 &1.44 &-1.61 &1.93 & 83 &-1.614 &0.095 &17 &-1.475 &0.098 \\ 
t01348 & 4473 &1.57 &-1.57 &1.52 & 74 &-1.571 &0.101 &15 &-1.586 &0.109 \\ 
t01554 & 4634 &1.76 &-1.63 &0.98 & 52 &-1.629 &0.140 &10 &-1.588 &0.173 \\ 
t01699 & 4546 &1.72 &-1.52 &1.64 & 76 &-1.517 &0.090 &15 &-1.517 &0.105 \\ 
\multicolumn{11}{c}{GIRAFFE} \\
t00261 & 3922 &0.51 &-1.56 &2.09 & 40 &-1.558 &0.087 & 3 &-1.524 &0.207 \\ 
t00284 & 4022 &0.71 &-1.59 &1.97 & 36 &-1.592 &0.099 & 3 &-1.517 &0.064 \\ 
t00316 & 4056 &0.77 &-1.58 &2.03 & 48 &-1.576 &0.123 & 3 &-1.576 &0.216 \\ 
t00329 & 3968 &0.59 &-1.59 &2.02 & 38 &-1.588 &0.079 & 3 &-1.514 &0.089 \\ 
t00337 & 3971 &0.59 &-1.61 &2.08 & 41 &-1.605 &0.110 & 3 &-1.512 &0.016 \\ 
\hline
\end{tabular}
\label{tabatmo}
\end{table*}

\setcounter{table}{4}
\begin{table*}
\centering
\caption[]{Light element abundances for confirmed member stars.
{\bf the complete table is available in the on-line version}}
\begin{tabular}{rccccrcccrcccrcccrcccrcccrccclc}
\hline
Star   &   nr &  [O/Fe]{\sc i} &  $rms$ &  nr &  [Na/Fe]{\sc i} &  $rms$ &  nr &  [Mg/Fe]{\sc i} &  $rms$ &  nr &  [Al/Fe]{\sc i} &  $rms$ &Lim? &PIE \\
\hline
\multicolumn{14}{c}{UVES} \\
t00540 &  1 &-0.041 &	    &3 & 0.637 &0.010    & 3  &0.518 &0.068 &2 & 0.802 &0.008  &     &  I\\ 
t00553 &  2 & 0.174 &0.071  &3 & 0.303 &0.050    & 3  &0.459 &0.038 &2 & 0.395 &0.084  &     &  I\\ 
t00670 &  1 &-0.126 &	    &2 & 0.377 &0.089    & 2  &0.420 &0.071 &2 & 0.559 &0.006  &     &  I\\ 
t00951 &  1 & 0.506 &	    &1 &-0.028 &         & 2  &0.482 &0.011 &2 & 0.330 &0.126  &     &  P\\ 
t01348 &  1 & 0.369 &	    &2 & 0.030 &0.055    & 2  &0.464 &0.006 &1 & 0.270 &       &     &  P\\ 
t01554 &  1 & 0.263 &	    &2 & 0.313 &0.061    & 2  &0.532 &0.204 &2 & 0.724 &0.097  &     &  I\\ 
t01699 &  1 & 0.146 &	    &2 & 0.414 &0.001    & 2  &0.514 &0.102 &2 & 0.536 &0.112  &     &  I\\ 
\multicolumn{14}{c}{GIRAFFE} \\
t00261 &  2 & 0.078 &0.185  &4 & 0.631 &0.040    & 3  &0.503 &0.072 &2 & 0.555 &0.031  &     &  I\\ 
t00284 &  2 & 0.209 &0.006  &4 & 0.657 &0.040    & 3  &0.563 &0.057 &2 & 0.556 &0.000  &     &  I\\ 
t00316 &  1 & 0.037 &	    &4 & 0.637 &0.061    & 2  &0.503 &0.006 &2 & 0.940 &0.128  &     &  I\\ 
t00329 &  2 & 0.430 &0.051  &4 & 0.340 &0.036    & 3  &0.538 &0.015 &2 & 0.279 &0.006  &     &  I\\ 
t00337 &  2 & 0.105 &0.006  &4 & 0.747 &0.019    & 3  &0.516 &0.076 &2 & 0.921 &0.040  &     &  I\\ 
\hline
\end{tabular}
\label{tablight}
\end{table*}

\end{document}